\documentclass{iaus}
\usepackage[latin1]{inputenc}  
\usepackage{graphicx}

\title[Destruction of gaseous biomolecules by VUV and soft X-rays] 
{Survival of gas phase amino acids and nucleobases in space radiation
conditions}

\author[Pilling et al.]   
{S. Pilling$^1$, D. P. P. Andrade$^3$, R. B. de Castilho$^3$, R. L.
Cavasso-Filho$^1$, A. F. Lago$^1$, L. H. Coutinho$^2$, G. G. B. de Souza$^3$,
H. M. Boechat-Roberty$^3$ \and A. Naves de Brito$^{1}$}

\affiliation {$^1$LNLS, Laboratório Nacional de Luz Síncrotron, São Paulo,
Brazil.
\\ email: {\tt spilling@lnls.br, cavasso@lnls.br, alago@lnls.br, arndaldo@lnls.br} \\[\affilskip]
$^2$ UEZO, Centro Universitário Estadual da Zona Oeste, Rio de Janeiro,
Brazil. \\ email: {\tt coutinholh@yahoo.com} \\[\affilskip]
$^3$ UFRJ, Universidade Federal do Rio de Janeiro, Rio de Janeiro, Brazil.
email: {\tt diana\_andrade@ufrj.br, castilho@iq.ufrj.br, gerson@iq.ufrj.br,
heloisa@ov.ufrj.br}}

\pubyear{2008}
\volume{IAUS251}  
 \pagerange{? -- ?}
 \date{?? and in revised form ??}
 \jname{Organic Matter in Space} \editors{A.C. Editor,
B.D. Editor \& C.E. Editor, eds.}
\begin{document}

\maketitle

\begin{abstract}
We present experimental studies on the photoionization and photodissociation
processes (photodestruction) of gaseous amino acids and nucleobases in
interstellar and interpla\-netary radiation conditions analogs. The
measurements have been undertaken at the Brazilian Synchrotron Light Laboratory
(LNLS), employing vacuum ultraviolet (VUV) and soft X-ray photons. The
experimental set up basically consists of a time-of-flight mass spectrometer
kept under high vacuum conditions. Mass spectra were obtained using
photoelectron photoion coincidence technique. We have shown that the amino
acids are effectively more destroyed (up to 70-80\%) by the stellar radiation
than the nucleobases, mainly in the VUV. Since polycyclic aromatic hydrocarbons
have the same survival capability and seem to be ubiquitous in the ISM, it is
not unreasonable to predict that nucleobases could survive in the interstellar
medium and/or in comets, even as a stable cation.

\end{abstract}
\firstsection 
%
\section{Introduction}

The search for amino acids and nucleobases (and related compounds) in the
interstellar medium/comets has been performed at least in the last 30 years,
but unfortunately it was not successful yet (e.g. Brown et al. 1979; Simon \&
Simon, 1973). However, recently some traces (upper limits) of simplest amino
acid, glycine (NH$_2$CH$_2$COOH) were observed in the comet Hale-Bopp
(Crovisier et al. 2004) and in some molecular clouds associated with the star
forming regions (Kuan et al. 2003a) but these identifications have yet to be
verified (Snyder et al. 2005; Cunningham et al. 2007). Despite no direct
detection of nucleobases in comets or in the molecular clouds, some of their
precursors molecules like HCN, pyridines, pyrimidines and imidazole were been
reported in the Vega 1 flyby of comet Halley (Kissel \& Krueger, 1987) and have
been searched in the interstellar medium (Kuan et al. 2003b).

The search for these biomolecules in meteorites, on the contrary, has been
revealed an amazing number of proteinaceous and non-proteinaceous amino acids,
up to 3 parts per million (ppm) (e.g. Cronin 1998 and references therein), and
some purine and pyrimidine based nucleobases up to 1.3 ppm (e.g. Stocks \&
Schwartz 1981 and references therein). This dichotomy between the carbonaceous
chondrites meteorites and interstellar medium/comets chemistry remains a big
puzzle in astrochemistry filed and in the investigation about the origin of
life.

The goal of this work is to review some experimental gas-phase photoionization
and photodissociation studies of amino acids and nucleobases induced by vacuum
ultra-violet (VUV) and soft X-ray photons, obtained recently by our group (Lago
et al. 2004; Coutinho et al. 2005; Marinho et al. 2006; Pilling et al 2007c). A
possible direction on the different survivability of these biomolecules on
astrophysical environments are given.

\section{Experimental methodology and results}
In an attempt to simulate the stellar/solar VUV and soft X-ray flux we have
used a synchrotron radiation as a light source. The experiments were performed
at the Brazilian Synchrotron Light Laboratory (LNLS), employing harmonic free
VUV photons (Cavasso-Filho et al. 2007) and soft X-ray photons from the
toroidal grating monochromator (TGM) beamline. The emergent photon beam flux
was recorded by a light sensitive diode. Briefly, the radiation
($\sim$10$^{12}$ photons s$^{-1}$) from the beamline perpendicularly intersect
the vapor-phase sample at the center of the ionization region inside a high
vacuum chamber (Boechat-Roberty et al. 2005; Pilling et al. 2006).  Mass
spectra were obtained using photoelectron photoion coincidence (PEPICO)
technique (Pilling et al 2007a; 2007b and references therein).

\begin{figure}[!bt]
\begin{center}
 \includegraphics[width=5in]{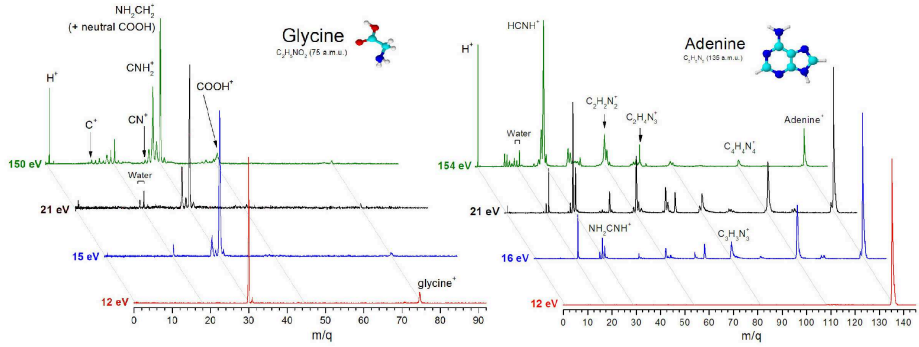}
\caption{\scriptsize{Time-of-flight mass spectra of gaseous amino acid glycine
and nucleobase adenine recorded at different photon energies at VUV (12-21 eV)
and at soft X-ray ($\sim$ 150 eV) ranges.}}
   \label{fig-1}
\end{center}
\end{figure}

In Fig.~\ref{fig-1} we have shown the time-of-flight mass spectra of the
fragments released from the amino acid glycine and the nucleobase adenine,
recorded at different photon energies over the VUV (12-21 eV) and soft X-ray
($\sim$ 150 eV) ranges. As a general rule, even at low photon energy, the amino
acids have only a small contribution ($\lesssim$ 10\%) or even they were not
detected, a consequence of their high photodestruction degree. The release of
the carboxyl group (COOH) as a neutral or cationic species, depending on the
amino acid, is one of the most import dissociation channels (see also Jochims
et al. 2004). The nucleobases have shown a higher molecular stability in
comparison with the amino acids ones. For these molecules, the parent ions
remain relatively strong over all the studied VUV photon energy range. In all
spectra, as the photon energy increases also increase the fragmentation
profile, as expected. Some minor contamination of water was observed in the
spectra, which reflects the high hydrophilic character of the samples.

In the previous studies of photodissociation of nucleobases (e.g. Jochims et
al. 2005; Schwell et al. 2006) at the VUV photon energy range the authors have
identified some important photodissociation channels as well the HNCO loss by
thymine and uracil and HCN loss by adenine. As pointed by Pilling et al.
(2007c), in the case o adenine, the neutral HCN may represent as much as 40\%
of its photodissociative channels. The ion HCNH$^+$ is another largely
photoproduced fragment from both amino acids and nucleobases.

\begin{figure}[!tb]
\begin{center}
 \includegraphics[width=3.3in]{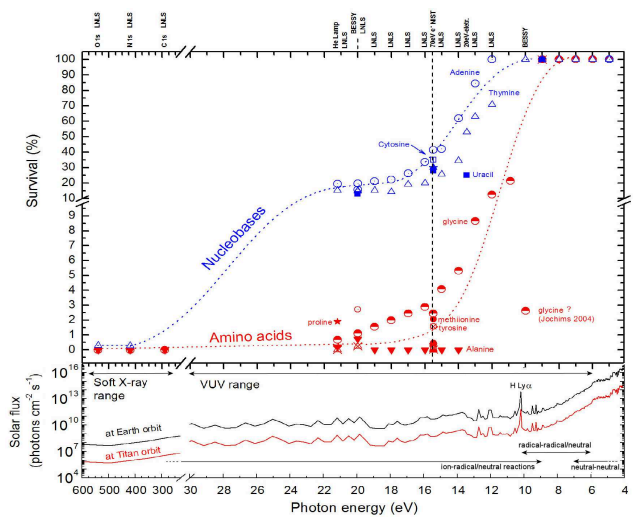}
 \caption{\scriptsize{Comparison between the
survival (photoresistence) of different amino acids and nucleobases due to
ionizing radiation field in the VUV and soft X-ray. The solar photon flux at
Earth and Titan orbit (adapted from Gueymard 2004). See details in text.}}
   \label{fig-2}
\end{center}
\end{figure}

A comparison between the survival of different nucleobases and amino acids due
to ionizing radiation field from 4 to 600 eV is given at Fig.~\ref{fig-2}. Our
data is represented by LNLS labels. The literature data were gather mainly from
the NIST database and Jochims et al. (2004; 2005). The amino acids are
effectively more destroyed by VUV stellar/solar radiation than the nucleobases.
The higher resistance of nucleobases to the ionizing photons may be associated
with the presence of the hetero-cyclic structure and unsaturated bonds.

For comparison, we also present in the bottom panel of Fig.~\ref{fig-2}, the
solar photon flux at ultraviolet and soft X-rays at Earth and Titan orbit
(adapted from Gueymard 2004). The different photochemical domains are also
given. According to the detailed review of Schwell et al. (2006), the
photoabsorption cross section of these molecules is higher in VUV as compared
to the mid-UV ($<$ 6 eV). In particular all molecules, absorb strongly at 10.2
eV, where the intense Ly$\alpha$ (10.2 eV) stellar/solar emission is located.
Most of the small biomolecules studied have first ionization energies (IE)
below this energy, making photoionization phenomena an important issue to
study. Since most of the amino acids has the first AE below 10.2 eV, the
stellar/solar hydrogen Ly$\alpha$ has a great influence on their molecular
survival. This is not observed in the case of nucleobases which the first AE
occurs at energy above the hydrogen Ly$\alpha$, where the photon flux is about
2 orders of magnitude lower (in the case of Sun).

\section{Conclusion}
We have shown that the amino acids are effectively more destroyed by stellar
radiation than the nucleobases, mainly at VUV spectral range where the
differences reach up 70-80 \% decreasing to high-energy photon range,
corroborating other experimental results given in the literature. The
nucleobases are able to form a stable cation in gas phase and since polycyclic
aromatic hydrocarbons (PAHs) and polycyclic aromatic nitrogen-rich hydrocarbons
(PANHs) have the same capability and seem to be ubiquitous in the ISM
(Allamandola et al. 1986), it is not unreasonable to predict that aromatic
nucleic acid bases could survive in the interplanetary and interstellar medium.

These results lead us to make an interesting question. Why we did not find
nucleobases in cometary/molecular clouds radioastronomical observation since
they are quite more resistant to stellar ultraviolet radiation than the
detected amino acid (e.g. glycine)? Probably, the answer may be associated with
the formation pathways efficiencies rather than with the detection limits and
more studies over this subject are need.

Finally, a possible direction to the search of large pre-biotic molecules, as
the case of amino acids and nucleobases or even for larger molecules, could be
the search not for the molecules themselves, but from their photoproduced
daughter species like, for example, the fragments COOH$^+$, HNCO$^+$, HCN$^+$
and HCNH$^+$. The abundances, derived from the radio-observation of these
fragments, combined with the laboratory data (e.g. relative ion yield and
photoproduction cross section) may give us a clue about the presence and the
amount (upper limit) of their parent molecules.


\end{document}